\documentclass[runningheads]{llncs}
\usepackage{graphicx}
\usepackage{diagbox}
\usepackage{xcolor}
\usepackage{todonotes}

\begin{document}
\title{Deep Learning-based Segmentation of Pleural Effusion From Ultrasound Using Coordinate Convolutions}
\titlerunning{Deep Learning-based Segmentation of Pleural Effusion}

\author{Germain Morilhat\inst{2,1} \and Naomi Kifle \inst{1,3} \and Sandra FinesilverSmith \inst{4} \and Bram Ruijsink \inst{1,5} \and Vittoria Vergani\inst{6} \and Habtamu Tegegne Desita \inst{7} \and Zerubabel Tegegne Desita \inst{8} \and Esther Puyol-Ant\'{o}n \inst{1} \and Aaron Carass \inst{3} \and Andrew P. King \inst{1}}

\authorrunning{G. Morilhat et al}

%
\institute{School of Biomedical Engineering \& Imaging Sciences, King's College London, U.K. \and IMT Atlantique, France \and Image Analysis and Communications Laboratory,
Johns Hopkins University, U.S.A. \and Big Data Institute, Oxford University, U.K. \and Department of Cardiology,
University Medical Center Utrecht, The Netherlands. \and King's College Hospital, London, U.K. \and Department of Information Technology, University of Gondar, Ethiopia. \and Department of Radiology, University of Gondar, Ethiopia. }

\maketitle              % typeset the header of the contribution
\begin{abstract}
Ultrasound imaging plays a crucial role in assessing disease and making diagnoses for a range of conditions, especially so in low-to-middle-income (LMIC) countries. One such application is the assessment of pleural effusion, which can be associated with multiple morbidities including tuberculosis (TB). Currently, assessment of pleural effusion is performed manually by the sonographer during the ultrasound examination, leading to significant intra-/inter-observer variability. In this work, we investigate the use of deep learning (DL) to automate the process of pleural effusion segmentation from ultrasound images. On two ultrasound datasets of suspected TB patients acquired in a LMIC setting, we achieve median Dice Similarity Coefficients (DSCs) of 0.82 and 0.74 respectively using the nnU-net DL model. We also investigate the use of coordinate convolutions in the DL model and find that this results in a statistically significant improvement in the median DSC on the first dataset to 0.85, with no significant change on the second dataset. This work showcases, for the first time, the potential of DL in automating the process of effusion assessment from ultrasound imaging and paves the way for future work on artificial intelligence-assisted acquisition and interpretation of ultrasound images. This could enable accurate and robust assessment of pleural effusion in LMIC settings where there is often a lack of experienced radiologists to perform such assessments.

\end{abstract}

\section{Introduction}

In 2020 there were an estimated 10 million cases of tuberculosis (TB) worldwide, and the global case fatality ratio was 15\% \cite{WHO2021}. However, the prognosis for TB is generally good if treatment can be initiated early enough. The gold standard for diagnosis of TB is detection of mycobacterium tuberculosis through a culture test. However, such tests can be expensive and time-consuming, limiting their utility in low-to-middle-income (LMIC) countries \cite{Parsons2011}. In many LMIC countries radiological indicators play an important role in assessing clinical symptoms associated with TB, with a view to initiating treatment. One such symptom is pleural effusion, which refers to a build-up of excess fluid between the layers of the pleura outside the lungs. It can be caused by TB and several other conditions, including congestive heart failure, kidney failure, cancer, pneumonia, and pulmonary embolism. Pleural effusion can be identified using a chest X-ray but the sensitivity of this method is only good when the effusion volume is large, which makes it unsuitable for initiating early treatment.

Ultrasound imaging allows earlier identification of pleural effusion and grading of its severity, allowing better treatment allocation \cite{Soni2015}. However, ultrasound image 
%Ultrasound imaging has the potential to identify pleural effusion earlier, but
acquisition and interpretation for pleural effusion assessment require expertise, and in many LMIC countries there is a shortage of skilled sonographers who can perform this task.
Even for experienced sonographers, ultrasound-based assessment of pleural effusion is a challenging task, made difficult by the fact that the lungs can change appearance in ultrasound images in the presence of some pathologies, and the appearance of the effusion itself can change as it progresses from exudate to empyema.
Furthermore, shadowing caused by the ribs in ultrasound imaging can make it difficult to reliably measure the extent of the effusion.
As well as identifying the presence of effusion it is also useful to know its severity. Typically, effusion severity is estimated by manually measuring the ``depth'' of the effusion in ultrasound (i.e. the perpendicular distance between the pleural boundary and the lung). However, this measurement will vary depending on the probe orientation and how superior/inferior the measurement is taken, and so there is significant inter-/intra-observer variability.

In other applications, deep learning techniques have been used to automate medical image analysis tasks with a view to reducing intra-/inter-observer variability. For example, in ultrasound, deep learning has been applied to cardiac functional quantification \cite{Ouyang2020,Tromp2021}, assessing kidney function \cite{Kuo2019} and estimating fetal biometrics \cite{Zhang2020}. Across a range of recent medical image segmentation challenges, the nnU-Net framework \cite{Isensee2021} has proved to exhibit state-of-the-art performance.

In this paper we investigate the potential of deep learning to automate the task of pleural effusion segmentation from ultrasound imaging. To the best of our knowledge this is the first attempt to automate this challenging task.
We employ the state-of-the-art nnU-Net framework and also investigate whether coordinate convolutions can improve performance by explicitly encoding spatial information to improve the model's learning. Coordinate convolutions were first proposed in \cite{Liu2018} and have since been shown to improve performance or optimisation properties in medical image segmentation tasks \cite{ElJurdi2020}. In our application, due to the standard protocols used for acquiring effusion images, there is good reason to suspect that spatial information may improve segmentation performance and we investigate this hypothesis in this paper.
Some previous works have demonstrated this potential in ultrasound image analysis tasks \cite{Youn2020,Saleh2021,Wang2020} and here we investigate its potential for the task of pleural effusion segmentation in suspected TB patients.

The primary goal of this work is to demonstrate
the feasibility of deep learning methods to tackle the pleural effusion segmentation task. Our first contribution is using the
state-of-the-art nnU-Net deep learning model to address this task. Our second contribution is to investigate the use of spatial context information by extending the nnU-net model to use coordinate convolutions.

\section{Materials}

All ultrasound images were acquired using a SONOACE X7 ultrasound machine by an experienced radiologist at Gondar University Hospital in Ethiopia. Patients underwent clinical examination after reporting with symptoms consistent with a possible diagnosis of TB. All gave informed consent to the use of their images for research purposes and the study was approved by the university’s hospital administration. All images were stored in DICOM format and pseudonymised (including blanking of patient details in the image) before being transferred to a password-protected remote file server for subsequent analysis.

\begin{figure*}
\begin{minipage}[t]{0.365\textwidth}
\centering
  \includegraphics[width=0.97\linewidth]{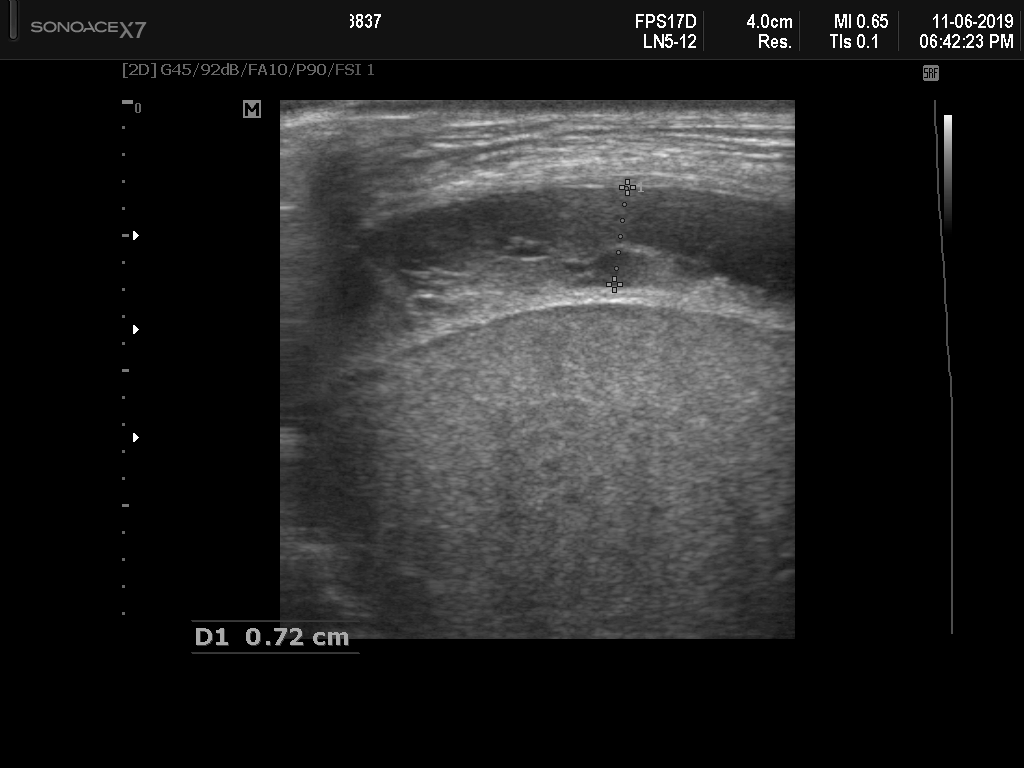}
\end{minipage}%
\hfill % maximize the horizontal separation
\begin{minipage}[t]{0.365\textwidth}
\centering
  \includegraphics[width=0.97\linewidth]{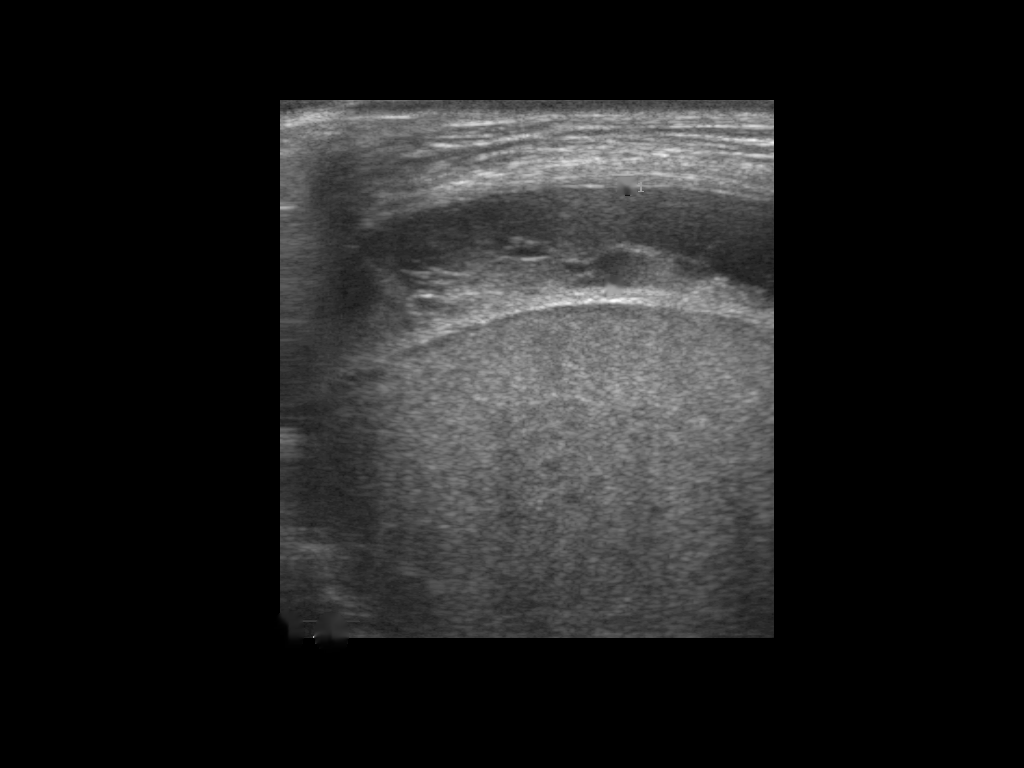}
\end{minipage}%
\hfill % maximize the horizontal separation
\begin{minipage}[t]{0.25\textwidth}
\centering
  \includegraphics[width=0.97\linewidth]{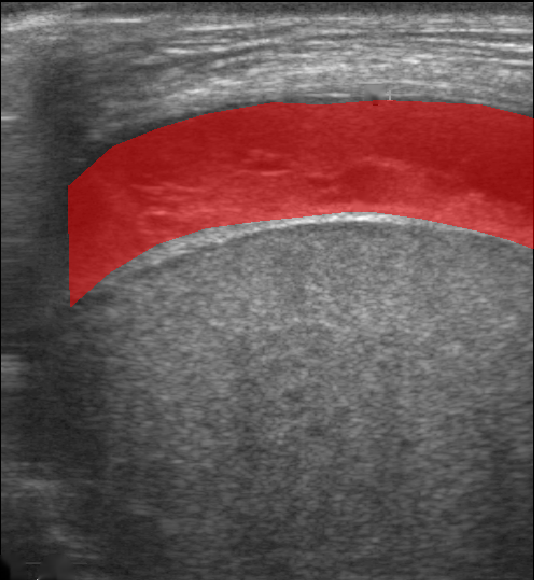}
\end{minipage}%
~\\
~\\
\begin{minipage}[t]{0.335\textwidth}
\centering
  \includegraphics[width=0.97\linewidth]{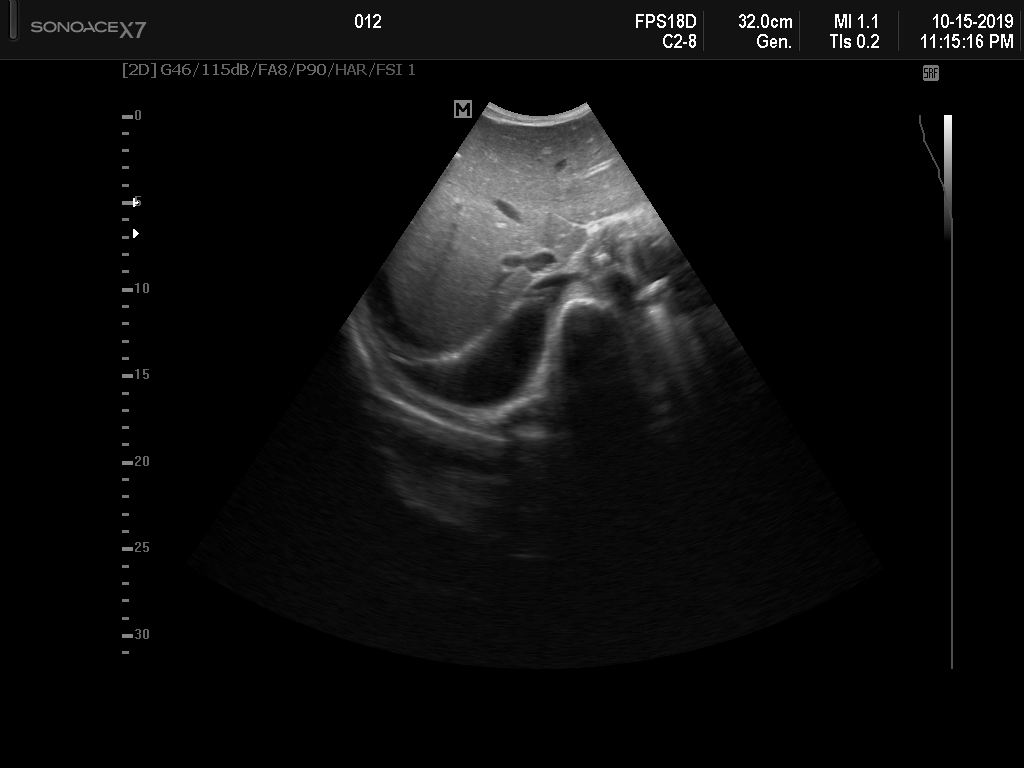}
\end{minipage}%
\hfill % maximize the horizontal separation
\begin{minipage}[t]{0.335\textwidth}
\centering
  \includegraphics[width=0.97\linewidth]{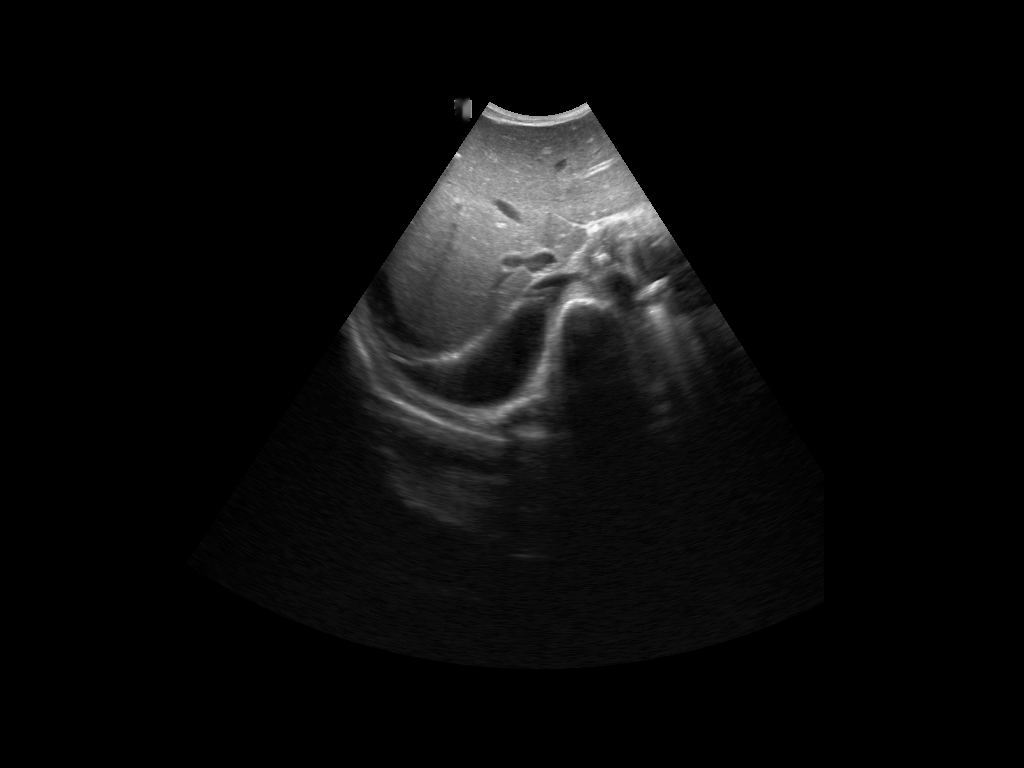}
\end{minipage}%
\hfill % maximize the horizontal separation
\begin{minipage}[t]{0.31\textwidth}
\centering
  \includegraphics[width=0.97\linewidth]{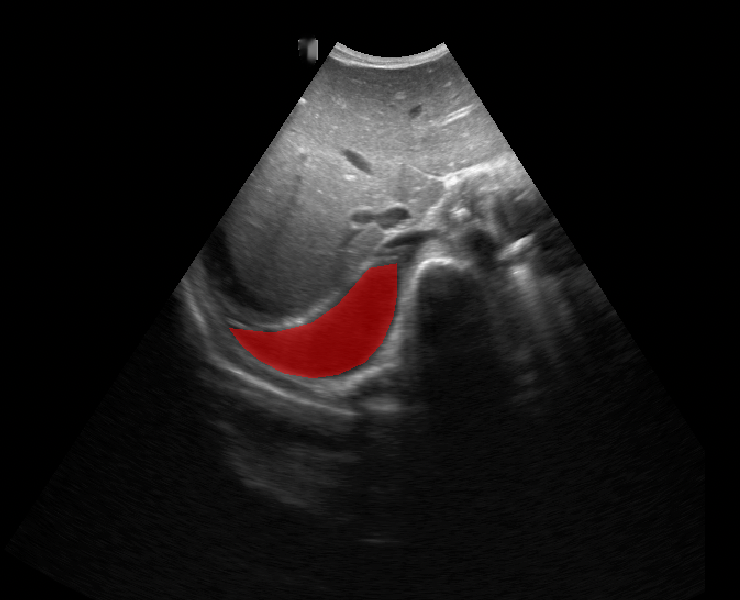}
\end{minipage}%
\caption{Sample ultrasound images. Left-to-right: original image, image after cropping and inpainting to remove annotations, further cropping, with ground truth segmentation overlaid in red. Top row: Dataset A. Bottom row: Dataset B.}
\label{fig:images}
\end{figure*}

A total of 143 images were acquired from 59 patients. The images were obtained at the left and right PLAPS (PosteroLateral Alveolar and/or Pleural Syndrome) and subcostal views \cite{Lichtenstein2011} with linear array and curved array (abdominal) ultrasound probes. The data were split according to the use of these probes into two datasets of 51 and 92 images, respectively for linear array and curved array. We denote these datasets as Dataset A (linear array) and Dataset B (curved array).
All images were annotated at the time of acquisition to measure the extent of the effusion. These annotations consisted of small crosses at the top and bottom of the deepest area of effusion.
See Figure \ref{fig:images} (left column) for example images.

Before being used for training and evaluating the models, each image was automatically cropped using a rectangular/cone mask to remove non-imaging content. Next, we applied an inpainting text algorithm using \emph{keras-ocr} followed by template matching and edge detection algorithms from \emph{opencv} to remove the annotations that were added to the images to measure the effusion. Examples of the outputs of this preprocessing are shown in Figure \ref{fig:images} (centre column).

All images in both datasets were manually segmented using the ITK-SNAP software \cite{Yushkevich2006} (\url{www.itksnap.org}) by a trained observer. Examples of ground truth segmentations are shown in Figure \ref{fig:images} (right column).
These segmentations acted as ground truths for training and evaluating the proposed models. Additionally, a second trained observer performed independent segmentations of subsets of 10 random images each from the two datasets. These were used to compute an estimate of inter-observer variability in the manual segmentation process.

\section{Methods}

For our baseline model we employed the nnU-Net deep learning framework \cite{Isensee2021}. We used the 2-D implementation and the model was trained for 100 epochs to limit computational demands. Training was performed  with a batch size of 4 using stochastic gradient descent with Nesterov momentum ($\mu$=0.99) and an initial learning rate of 0.01. The loss function was the sum of cross entropy and Dice loss. The default nnU-Net data augmentation setting was used which included rotations, scaling, Gaussian noise, Gaussian blur, brightness, contrast, simulation of low resolution, gamma correction and mirroring. The model used for inference was the final model after all training epochs. We chose to use this model rather than the best model over the training epochs so that our results could be treated as test rather than validation results (see Section \ref{sect:expts}).

We also investigated whether using coordinate convolutions \cite{Liu2018} could improve the performance of the nnU-Net baseline. Coordinate convolutions work by adding extra channels to the input layer which contain the coordinates of the pixels. In our case, as our images are 2-D this involved adding two extra channels, one containing the $x$-coordinates and one containing the $y$-coordinates, with the coordinates being specified in pixels and the origin being at the top left of the image.

\section{Experiments}
\label{sect:expts}

%\todo[inline]{AK: Was the cross validation split done by patient or by image?

%Germain: The split was made amoung all the image, independently from patients. Images have really been treated like they were each other independant, which can maybe have implied a bias that can be reduce in the future. 
%}

Evaluation for both experiments was performed using a 5-fold cross validation, i.e. the data were split into 5 folds and each was held out in turn and evaluated on a model trained using the other 4 folds. Due to the large variation in appearance between images acquired from the same patients, we performed the cross validation split at the image level rather than the patient level. Because nnU-net performs hyperparameter optimisation using heuristic rules and not using the validation data, and we used the model from the last epoch rather than using the validation data for model selection, these cross validation results can be considered as independent test results.

All models were evaluated using the Dice Similarity Coefficient (DSC),
\begin{equation}
  \mbox{DSC} = \frac{2|X\cap Y|}{|X| +|Y|}
  \label{eq:DSC}
\end{equation}
where $X$ and $Y$ are the predicted and ground truth segmentations respectively. We report the median and lower/upper quartiles of DSC across the validation/test results of all images.

Additionally, we computed measures of the error and bias in estimation of effusion area, since these are likely to be clinically important measurements in effusion assessment. Specifically, we calculated:
%\begin{equation}
\begin{eqnarray}
    \mbox{Abs. area error \%} & = & (abs(|X| - |Y|) / |Y|) \times 100\% \\
    \mbox{Area bias \%} & = & ((|X| - |Y|) / |Y|) \times 100\%
\end{eqnarray}
%\end{equation}
where $|X|$ and $|Y|$ represent counts of the numbers of foreground pixels in the predicted and ground truth segmentations respectively. We report the median and lower/upper quartiles of these measures.

Finally, we also compute the DSC between the manual segmentations of the two observers on the subsets of 10 images for each dataset. The median DSCs are reported as estimates of inter-observer variability in manual segmentation.

\section{Results}

\begin{figure*}
\begin{minipage}[t]{0.3\textwidth}
\centering
  \includegraphics[width=0.9\linewidth]{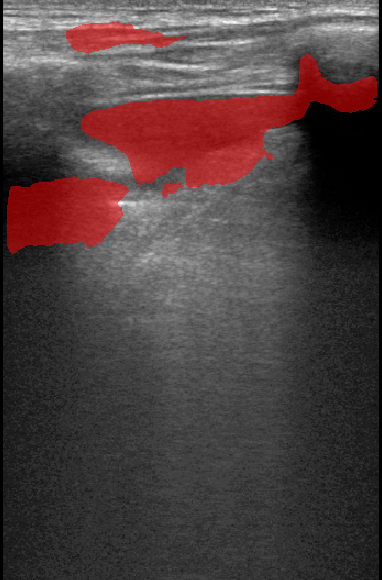}
\end{minipage}%
\hfill % maximize the horizontal separation
\begin{minipage}[t]{0.3\textwidth}
\centering
  \includegraphics[width=0.9\linewidth]{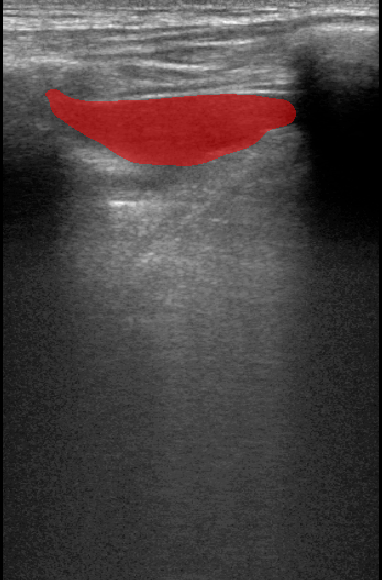}
\end{minipage}%
\hfill % maximize the horizontal separation
\begin{minipage}[t]{0.3\textwidth}
\centering
  \includegraphics[width=0.9\linewidth]{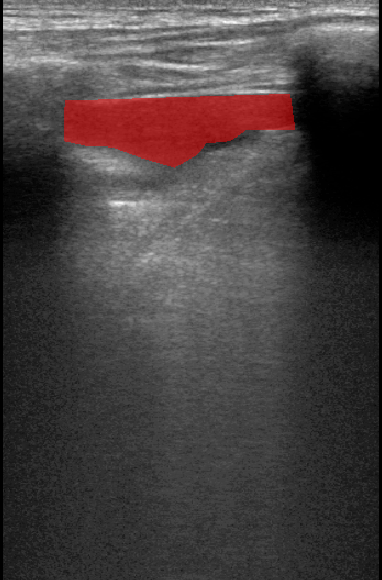}
\end{minipage}%
~\\
~\\
\begin{minipage}[t]{0.3\textwidth}
\centering
  \includegraphics[width=0.8\linewidth]{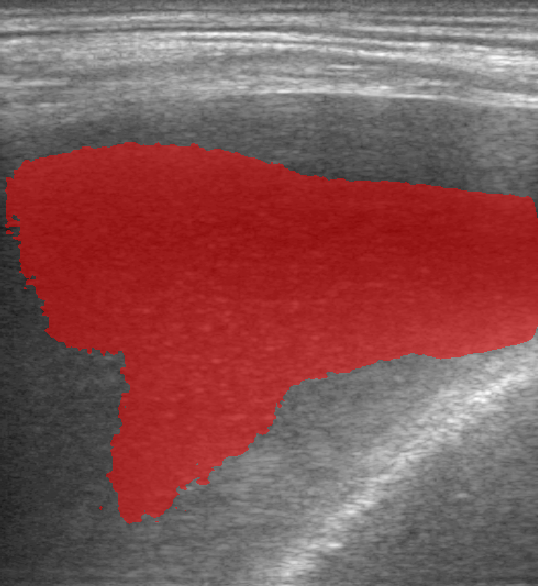}
\end{minipage}%
\hfill % maximize the horizontal separation
\begin{minipage}[t]{0.3\textwidth}
\centering
  \includegraphics[width=0.8\linewidth]{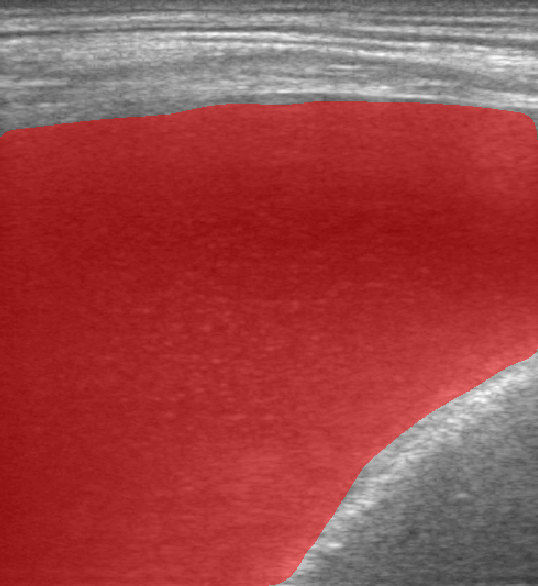}
\end{minipage}%
\hfill % maximize the horizontal separation
\begin{minipage}[t]{0.3\textwidth}
\centering
  \includegraphics[width=0.8\linewidth]{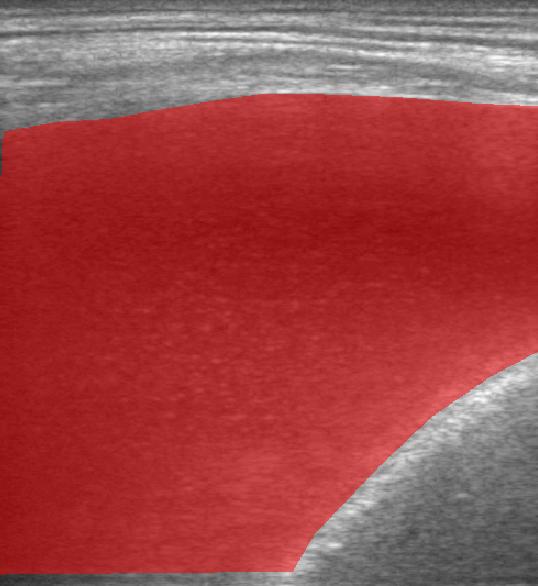}
\end{minipage}%
~\\
~\\
\begin{minipage}[t]{0.3\textwidth}
\centering
  \includegraphics[width=1\linewidth]{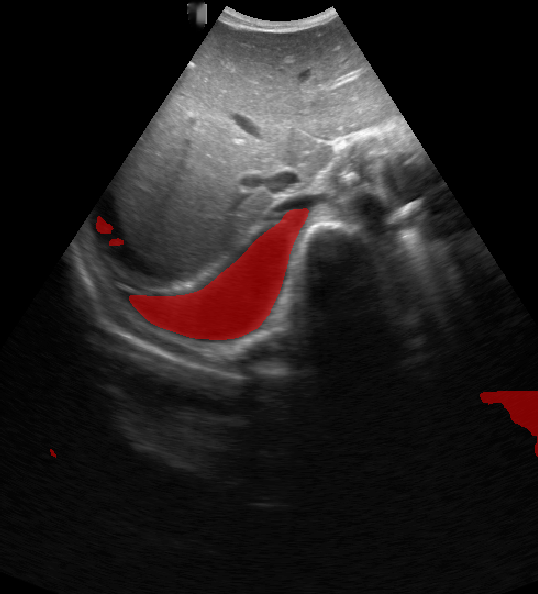}
\end{minipage}%
\hfill % maximize the horizontal separation
\begin{minipage}[t]{0.3\textwidth}
\centering
  \includegraphics[width=1\linewidth]{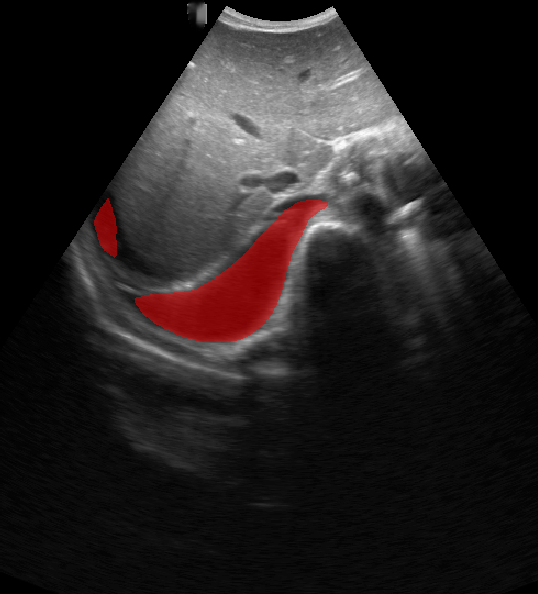}
\end{minipage}%
\hfill % maximize the horizontal separation
\begin{minipage}[t]{0.3\textwidth}
\centering
  \includegraphics[width=1\linewidth]{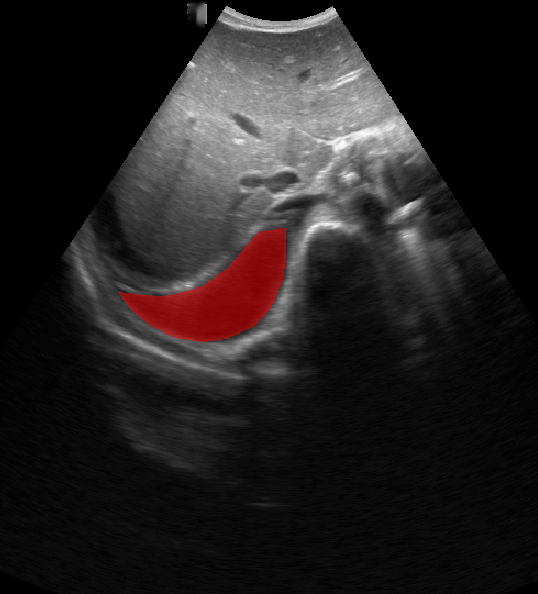}
\end{minipage}%
~\\
~\\
\begin{minipage}[t]{0.3\textwidth}
\centering
  \includegraphics[width=1\linewidth]{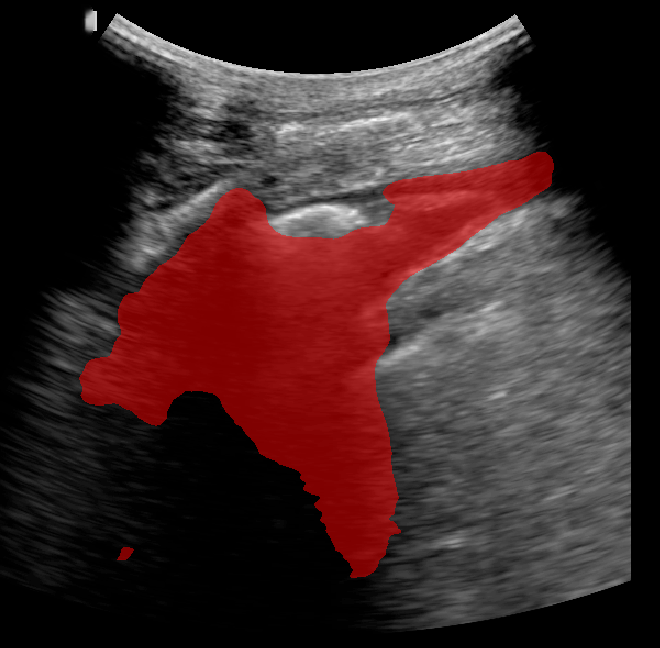}
\end{minipage}%
\hfill % maximize the horizontal separation
\begin{minipage}[t]{0.3\textwidth}
\centering
  \includegraphics[width=1\linewidth]{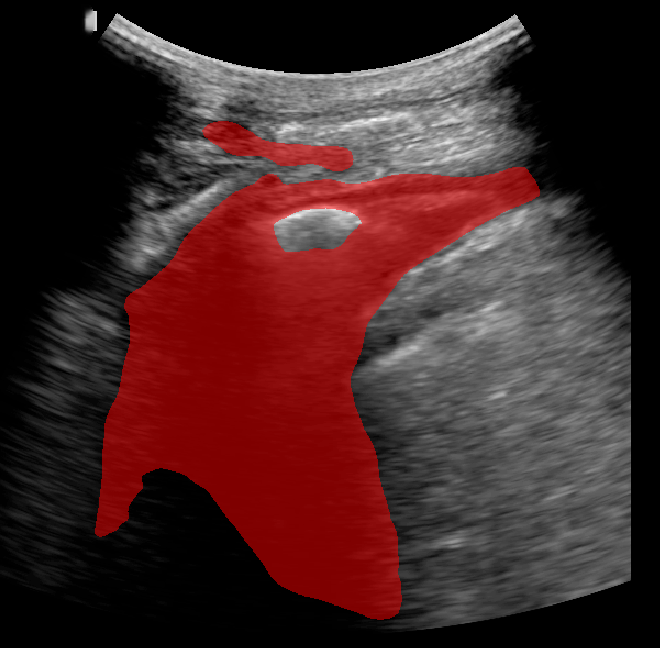}
\end{minipage}%
\hfill % maximize the horizontal separation
\begin{minipage}[t]{0.3\textwidth}
\centering
  \includegraphics[width=1\linewidth]{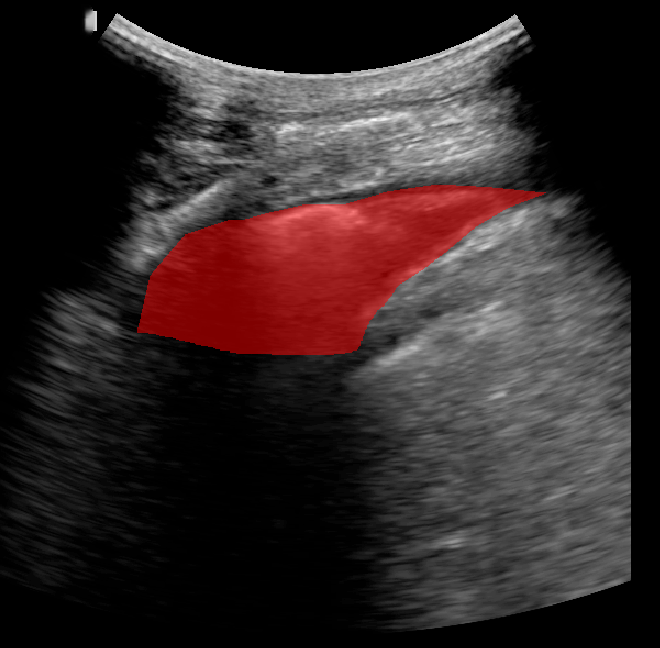}
\end{minipage}%
~\\
~\\
\caption{Model prediction results. Left-to-right: baseline model prediction, baseline with coordinate convolutions prediction, ground truth segmentation.\newline Rows 1-2: Dataset A. Rows 3-4: Dataset B.}
\label{fig:predictions}
\end{figure*}

\begin{figure*}
\begin{minipage}[t]{0.24\textwidth}
\centering
  \includegraphics[width=1.0\linewidth]{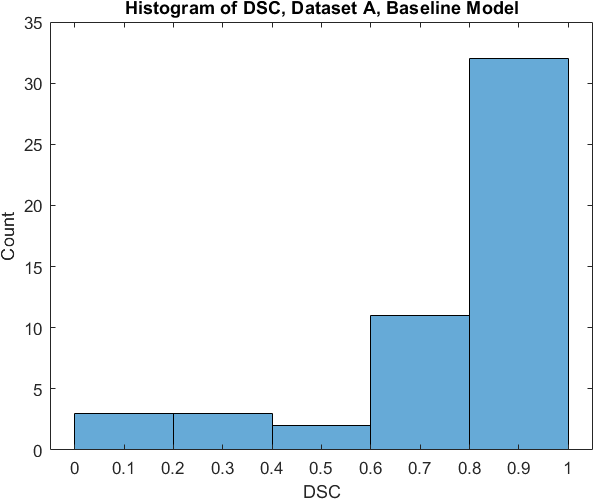}
\end{minipage}%
\hfill % maximize the horizontal separation
\begin{minipage}[t]{0.24\textwidth}
\centering
  \includegraphics[width=1.0\linewidth]{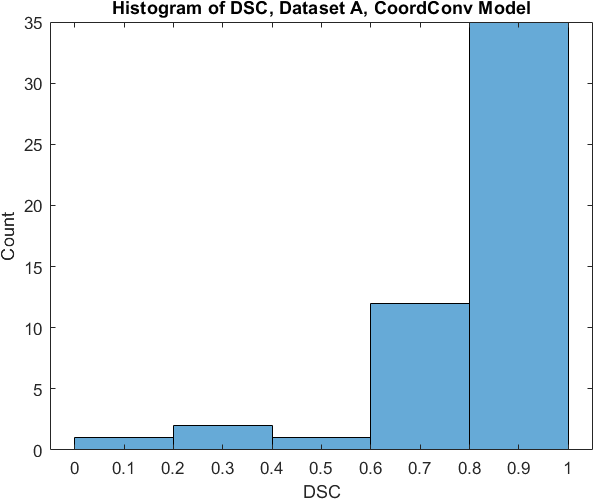}
\end{minipage}%
\begin{minipage}[t]{0.24\textwidth}
\centering
  \includegraphics[width=1.0\linewidth]{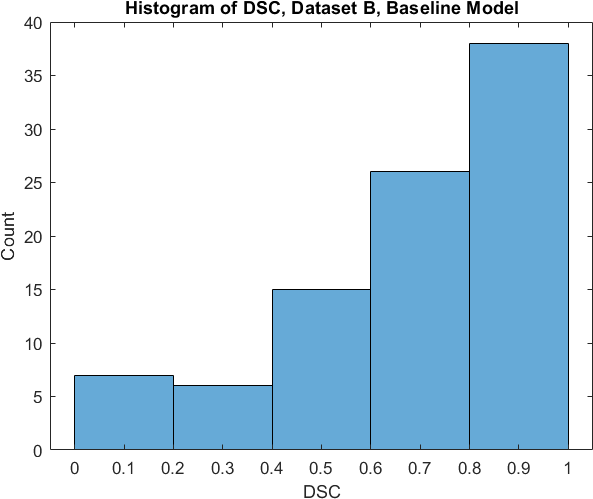}
\end{minipage}%
\hfill % maximize the horizontal separation
\begin{minipage}[t]{0.24\textwidth}
\centering
  \includegraphics[width=1.0\linewidth]{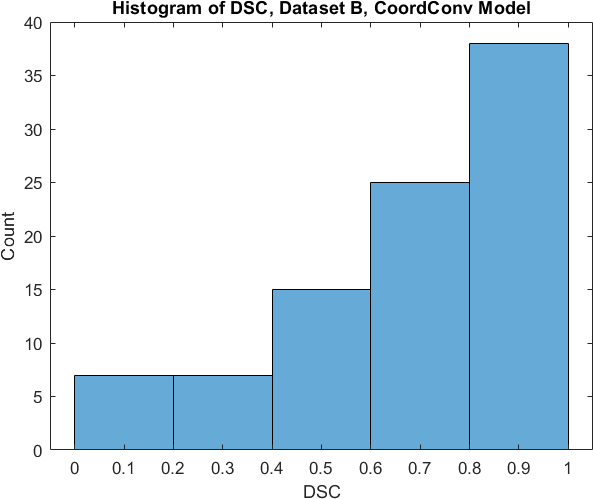}
\end{minipage}%
\caption{Histograms of DSC values, from left to right: baseline model (Dataset A), coordinate convolution model (Dataset A), baseline model (Dataset B), coordinate convolution model (Dataset B). Refer to Table \ref{table:results1} for summary statistics.}
\label{fig:histograms}
\end{figure*}

Qualitative prediction results of the two proposed models (baseline nnU-Net and nnU-Net with coordinate convolutions) on the two datasets are shown in Figure \ref{fig:predictions}.
%It can be seen that occasionally the baseline model introduces errors into the effusion segmentation but that the coordinate convolution model in general improves performance.
Tables \ref{table:results1} and \ref{table:results2} summarise the quantitative performances in terms of DSC and area statistics. Histograms of the DSC values are shown in Figure \ref{fig:histograms}. The median DSCs between the manual segmentations on the subsets of 10 images (i.e. the estimates of inter-observer variability) are also shown in Table \ref{table:results1}.

It can be seen that, despite having fewer images, the baseline model for Dataset A obtained a higher median DSC than the model for Dataset B.
For Dataset A the coordinate convolution model improved the DSC and reduced the area error and bias.
In two-tailed Wilcoxon signed rank tests (0.05 significance) the difference between the baseline DSC and that of the coordinate convolution model was found to be statistically significant for Dataset A ($p=0.0133$) but there was no statistically significant difference for Dataset B ($p=0.8$).
Interestingly, for both Dataset A and Dataset B, both the baseline and coordinate convolution models performed better than the estimate of inter-observer variability. However, we note that the inter-observer variability is quite high (i.e. median DSCs of 0.78 and 0.71), likely reflecting the difficulty and partly subjective nature of the effusion segmentation task. Therefore, it seems likely that the deep learning models are learning to segment effusion in the style of the main observer, which may not always be consistent with the second observer. In addition, the histograms shown in Figure \ref{fig:histograms} suggest that there are a significant number of failure cases in the outputs of both models (although fewer for the coordinate convolution model for Dataset A), again reflecting the difficulty of the task.

\begin{table}[ht]
\begin{center}
\caption{Summary of deep learning model segmentation performances. All figures are the median (lower, upper quartiles) of Dice Similarity Coefficients (DSC) across the entire dataset, computed using a 5-fold cross validation. For the inter-observer variability estimate, we quote only the median DSC. }
\begin{tabular}{c|ccc|}
\cline{2-4}
                       & \multicolumn{3}{c|}{\textbf{DSCs}}                            \\ \hline
\multicolumn{1}{|c|}{\textbf{Dataset}} & \multicolumn{1}{c|}{\textbf{Baseline}} & \multicolumn{1}{c|}{\textbf{Coord. conv.}} & \textbf{Inter-observer var.} \\ \hline
\multicolumn{1}{|c|}{A} & \multicolumn{1}{c|}{$0.82~(0.7, 0.89)$} & \multicolumn{1}{c|}{$0.85~(0.73, 0.92)$} & $0.78$ \\ \hline
\multicolumn{1}{|c|}{B} & \multicolumn{1}{c|}{$0.74~(057, 0.88)$} & \multicolumn{1}{c|}{$0.73~(0.55, 0.88)$} & $0.71$ \\ \hline
\end{tabular}
\label{table:results1}
\end{center}
\end{table}

%\begin{table}[ht]
%\begin{center}
%\begin{tabular}{l|ll|}
%\cline{1-3}
%\multicolumn{1}{|l|}{\textbf{Methods}} & \multicolumn{1}{l|}{\textbf{Abs. area error \%}} & \multicolumn{1}{l|}{\textbf{Area bias \%}}  \\ \hline
%\multicolumn{1}{|l|}{Baseline} & \multicolumn{1}{l|}{85 \%} & \multicolumn{1}{l|}{67 \%}   \\ \hline
%\multicolumn{1}{|l|}{Coord. conv.} & \multicolumn{1}{l|}{28 \% } & \multicolumn{1}{l|}{14 \%}  \\ \hline
%\end{tabular}
%\caption{Area statistics for dataset A }
%\label{table:results2}
%\end{center}
%\end{table}

\begin{table}[]
\begin{center}
\caption{Summary of deep learning model performamces in terms of area statistics. All figures are the median (lower, upper quartiles) across the entire dataset, computed using a 5-fold cross validation.}
\begin{tabular}{l|cc|cc|}
\cline{2-5}
                       & \multicolumn{2}{c|}{\textbf{Baseline}}    & \multicolumn{2}{c|}{\textbf{Coord. conv.}}    \\ \hline
\multicolumn{1}{|c|}{\textbf{Dataset}} & \multicolumn{1}{c|}{\textbf{Abs. area error \%}} & \textbf{Area bias \%} & \multicolumn{1}{c|}{\textbf{Abs. area error \%}} & \textbf{Area bias \%} \\ \hline
\multicolumn{1}{|c|}{A} & \multicolumn{1}{c|}{$20.0~(8.0, 63.4)$} & $1.9~(-17.6, 43.6)$  & \multicolumn{1}{c|}{$11.2~(4.7, 33.1)$} & $3.5~(-7.8, 19.7)$ \\ \hline
\multicolumn{1}{|c|}{B} & \multicolumn{1}{c|}{$19.9~(5.4, 60.0)$} & $1.5~(-16.1, 37.4)$ & \multicolumn{1}{c|}{$24.6~(7.9, 52.3)$} & $0.85~(-16.1, 33.4)$ \\ \hline
\end{tabular}
\label{table:results2}
\end{center}
\end{table}

%\todo[inline]{AK: Another suggestion has been to add a summary of results for effusion area. E.g. you could count the effusion pixels in the ground truths and predicted segmentations and for each image compute an absolute "error in area", perhaps expressed as a percentage of the ground truth area. We could include an extra table summarising these results like the DSC ones. Can you do this?}

%\todo[inline]{Germain: The issue is that sometimes the model has predicted more pixel than the ground truth really has so giving just a ratio of pixels doesn't seem to be really pertinant. Perhaps I should only true positive pixels...}

\section{Discussion and Conclusions}

To the best of our knowledge, we have presented the first study into the use of deep learning for automation of pleural effusion assessment from ultrasound images. Our results have demonstrated the potential of deep learning for this challenging task. The performance of the baseline model was superior to that of our reported inter-observer study, although we acknowledge that a number of failure cases remain (see e.g. the bottom row of Figure \ref{fig:predictions}). The coordinate convolution model improved performance for one of the two datasets (Dataset A, which was acquired using the linear array probe), but not for the other one. One possible explanation for this difference is that for the linear array probe one of the coordinates represents the distance from the probe (i.e. the $y$-coordinate). This may have made it easier for the model to exploit this potentially important piece of information. In future work we will examine more closely the impact of presenting spatial information to the model in different ways, e.g. using distance from the probe for the curved array probe dataset.
Nevertheless, this work represents an important proof-of-concept, paving the way for future work into artificial intelligence-assisted effusion assessment from ultrasound images.
Our eventual aim in this work is to reduce the need for skilled operators (who can be scarce in some LMIC settings) using machine learning techniques.

We have demonstrated the potential of our approach on a dataset of suspected TB patients. However, pleural effusion can be caused by a number of other factors and so we believe that our work will have wider potential applicability, both in LMIC settings and beyond. In fact, not all patients in our dataset were confirmed as TB cases. In addition to pleural effusion, pericardial effusion can also be assessed using ultrasound and we will investigate this possibility in future work.
In addition, different aspects of the effusion (which hold clues to the underlying disease process) could potentially be recognised using deep learning models, further reducing the need for skilled operators. 

One limitation of our work is the lack of control cases in our database. All of the subjects in the database had effusion identified clinically (although its severity was variable). Expansion of the dataset to include cases with no pleural effusion would enable a more robust model to be trained. Furthermore, it would be beneficial for our ground truths to be reviewed by a panel of trained observers to reach a consensus on where the effusion lies, to reduce uncertainty and variability in assessments between observers.

Our work has focused on the interpretation of ultrasound images, with a view to reducing intra-/inter-observer variability and (eventually) reducing the required skill level to widen access to ultrasound-based pleural effusion assessment in LMIC settings. However, in reality, acquiring good quality images of pleural effusion requires a certain level of skill. Therefore, the impact of our current work would be to speed up the workflows of skilled operators and to reduce intra-/inter-observer variability.
However, the standardised BLUE-protocol for lung ultrasound \cite{Lichtenstein2011} acquisition might require less skill compared to image interpretation.
Nevertheless, in the future, we will address the issue of image acquisition, and investigate the potential of machine learning techniques to simplify this process and enable less experienced operators to acquire good quality images. We envisage that this would involve some basic training and a simplified acquisition protocol combined with machine learning-based quality control and real-time integration of our automated effusion assessment model.

\section*{Acknowledgements}

This work was part-funded by a King’s College London Overseas Development Assistance Research Partnership Seed Fund award.

\bibliographystyle{splncs04}
\bibliography{references}
\end{document}